\documentclass[nobibnotes,nofootinbib,superscriptaddress,showkeys]{revtex4}
\usepackage{amsmath,amssymb,bm}
\usepackage{graphicx}

\newcommand{\beeq}{\begin{eqnarray}}
\newcommand{\eeeq}{\end{eqnarray}}
\newcommand{\be}{\begin{equation}}
\newcommand{\ee}{\end{equation}}
\newcommand{\bea}{\begin{array}}
\newcommand{\eea}{\end{array}}

\newcommand{\eq}{&=&}

\def\xp{x_{{I\!\!P}}}

\def\pbar{\overline{p}}
\def\qbar{\overline{q}}

\def\dbar{\overline{d}}
\def\ubar{\overline{u}}

\def\half{\textstyle{\frac{1}{2}}}

\def\funp{{I\!\!P}}

\begin{document}
\title{\bf Diffractive production of electroweak vector bosons at the LHC}

\author{Krzysztof Golec-Biernat}\email{golec@ifj.edu.pl}
\affiliation{Institute of Nuclear Physics Polish Academy of Sciences, Krak\'ow, Poland}
\affiliation{Institute of Physics, University of Rzesz\'ow, Rzesz\'ow, Poland}
\author{Agnieszka \L{}uszczak}\email{Agnieszka.Luszczak@ifj.edu.pl}
\affiliation{Institute of Nuclear Physics Polish Academy of Sciences, Krak\'ow, Poland}

\begin{abstract}
We analyse diffractive electroweak vector boson production in hadronic  collisions and show that 
the single diffractive $W$ boson
production asymmetry in rapidity is a particularly good observable at the LHC to test the concept of the flavour symmetric pomeron parton distributions. It may also provide an additional constraint for the parton distribution functions in the proton.
\end{abstract}

\maketitle

\section{Introduction}
\label{chapter:1}

The electroweak $W$ and $Z$ boson production in hadronic collision is a particularly valuable process 
to constrain parton distribution functions  (PDFs) in a nucleon. By measuring leptonic products of 
the weak boson decays, electroweak
parameters like $\sin^2\theta_W$, where $\theta_W$ is the effective weak mixing angle,  or the $W/Z$ boson masses and decay widths can also be determined.
At the Born level the $W$ and $Z$
bosons are produced from annihilation of two quarks in the colliding nucleons. 
In the collinear approximation, the elementary cross sections
for these processes have to be convoluted with the nucleons' PDFs. A direct access to these distributions
is provided by the measurement of $W^\pm$ production asymmetry  in rapidity. This quantity reflects
the fact that at given rapidity the two charged vector bosons are produced by quarks of different flavours.
The measured $W$ asymmetry can be used in the global fit analysis to constrain PDFs, 
in particular  the ratio of the $u$ and $d$  PDFs. Such measurements were done at the Fermilab Tevatron. The electron  charge asymmetry  in $W$ boson decays is presented in \cite{Abe:1994rj,Abazov:2008qv}, a direct 
measurement of the $W$ boson asymmetry is reported in \cite{Aaltonen:2009ta},
the forward-backward asymmetry of the electron  from $Z$ boson decays is discussed in \cite{Abe:1996us,:2008xq}, while  a short summary on the  $W$ nad $Z$  boson production at the Tevatron can be found 
in \cite{Nurse:2008xj}.

Diffractive hadroproduction of electroweak bosons was observed experimentally at the Tevatron  
\cite{Abe:1997jp} and analysed theoretically in a series of papers 
\cite{Covolan:1999sw,Covolan:2002kh,Covolan:2003ye,GayDucati:2007ps}. In a single diffractive dissociation case, one of the colliding hadrons remains intact while the other which dissociates into the diffractive state
is separated in rapidity from the intact hadron. In the pomeron model
interpretation of this process, the rapidity gap appears due to the exchange of  a pomeron, a vacuum quantum
number exchange, which in the case of diffractive processes with hard scale reveals partonic structure
\cite{Ingelman:1984ns}.
Thus,  the electroweak bosons are diffractively produced from the annihilation of two quarks, one from the
hadron and the other from the pomeron. The partonic structure of the pomeron is described by the pomeron parton distributions which are usually assumed to be flavour symmetric to account for vacuum quantum numbers of the pomeron. In the forthcoming analysis, we show that the measurement of the $W$ boson asymmetry in the 
diffractive $pp$ collisions at the LHC is an ideal process to test the pomeron model interpretation of the 
diffractive hadroproduction of electroweak bosons. In addition, this asymmetry may provide an additional constraint for the determination of the  PDFs in the proton.

The paper is organized as follows. In Section~\ref{chapter:2} we present basic formulas for the
weak boson production cross sections and discuss in detail the $W$ boson production
in the inclusive $p\pbar$ and $pp$ scattering. In Section~\ref{chapter:3} we analyse the diffractive
$W$ boson production at the LHC and present main results of our analysis.

\section{Hadroproduction of $W$ and $Z$ bosons}
\label{chapter:2}

\begin{figure}[t]
\begin{center}
\includegraphics[width=12cm]{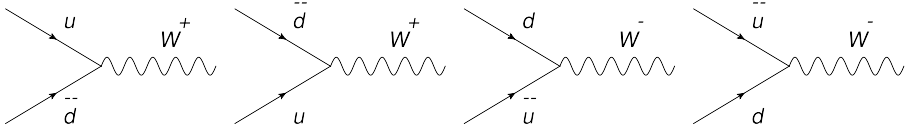}
\caption{The leading order diagrams for the $W^\pm$ boson production.}
\label{fig:0}
\end{center}
\end{figure}

The leading order cross sections for the subprocess $q\qbar^\prime\to W,Z$, with {\it collinear quarks},  are given by  \cite{Qcdandcollider}
\beeq\label{eq:3}
\sigma^{q\qbar^\prime\to W}\eq\frac{2\pi G_F}{3\sqrt{2}}M^2_W|V_{ff^\prime}|^2\delta(\hat{s}-M_W^2)
\\
\sigma^{q\qbar\to Z}\eq\frac{2\pi G_F}{3\sqrt{2}}M^2_Z(V_f^2+A_f^2)\delta(\hat{s}-M_Z^2)\,,
\eeeq
where $G_F$ is the Fermi constant and $V_{ff^\prime}$ is the appropriate CKM matrix element. In addition,
$V_f= T_f^3-2Q_f\sin^2\theta_W$
and
$A_f= T_f^3$
are the vector and axial couplings of the fermion $f$ to the $Z$ boson, respectively,  where
$T_f^3=\pm {\textstyle\half}$ with $(+)$ for the up type quarks and $(-)$ for the down type quarks  and $Q_f$ is given in units of the positron electric charge 
$e=g_w\sin\theta_W$.  These cross sections are convoluted
with the quark distribution functions taken at the scale, given by the corresponding vector boson mass, 
$\mu=M_{W,Z}$. Thus,   
we obtain for the $W$ bosons
\be\label{eq:4}
\frac{d\sigma_{W^{\pm}}}{dy}=\sigma_0^W\sum_{qq^\prime}|V_{qq^\prime}|^2\left\{
q(x_1,\mu)\,\qbar^\prime(x_2,\mu)\,+\,\qbar(x_1,\mu)\,q^\prime(x_2,\mu)\,,
\right\}
\ee
where the factorization scale $\mu=M_W$, and $q,\qbar$ denote quark/antiquark distributions. In addition
\be\label{eq:5}
\sigma_0^W=\frac{2\pi G_F}{3\sqrt{2}}\frac{M_W^2}{s}\,,~~~~~~~~~~
x_{1}=\frac{M_W}{\sqrt{s}} {\rm e}^{y}\,,~~~~~~~~~~
x_{2}=\frac{M_W}{\sqrt{s}} {\rm e}^{-y}\,.
\ee
where $y$ is   the W {\it boson rapidity}. Obviously, $y=\textstyle{\half}\ln(x_1/x_2)$ and from the condition $0<x_{1,2}<1$, the following constraint
results
\be
-y_{max}<y<y_{max}
\ee
with $y_{max}=\ln(\sqrt{s}/M_W)$.
The cross section for the $Z$ boson is obtained from Eq.~(\ref{eq:4}) by replacing
\be
|V_{qq^\prime}|^2~~ \to~~ \delta_{qq^\prime}(V_q^2+A_q^2)\equiv \delta_{qq^\prime}C_q\,.
\ee

In the forthcoming analysis we neglect the Cabbibo suppressed $s$ quark part of the $W$ production cross sections
and consider only two flavours: $u$ and $d$. Thus, for the partonic processes shown in Fig.~\ref{fig:0}, we find
\beeq
\frac{d\sigma_{W^+}}{dy}\eq \sigma_0^W\,|V_{ud}|^2\left\{u(x_1)\,\dbar(x_2)\,+\,\dbar(x_1)\,u(x_2)  \right\}
\\
\frac{d\sigma_{W^-}}{dy}\eq \sigma_0^W\,|V_{ud}|^2\left\{ d(x_1)\,\ubar(x_2)\,+\,\ubar(x_1)\,d(x_2)      \right\}
\\
\frac{d\sigma_{Z}}{dy}\eq \sigma_0^Z\,\left\{C_u\, u(x_1)\,\ubar(x_2)\,+\,C_d\,d(x_1)\,\dbar(x_2) \,+\,
(x_1\leftrightarrow x_2)     \right\},
\eeeq
where the parton distributions are taken at the scale $\mu=M_{W,Z}$.
The $W^{\pm}$ boson production asymmetry in rapidity is defined as follows
\be\label{eq:6}
A(y)=\frac{d\sigma_{W^+}(y)/dy-d\sigma_{W^-}(y)/dy}{d\sigma_{W^+}(y)/dy+d\sigma_{W^-}(y)/dy}\,.
\ee

\subsection{$p\bar{p}$ collisions}

\begin{figure}[t]
\begin{center}
\includegraphics[width=6cm]{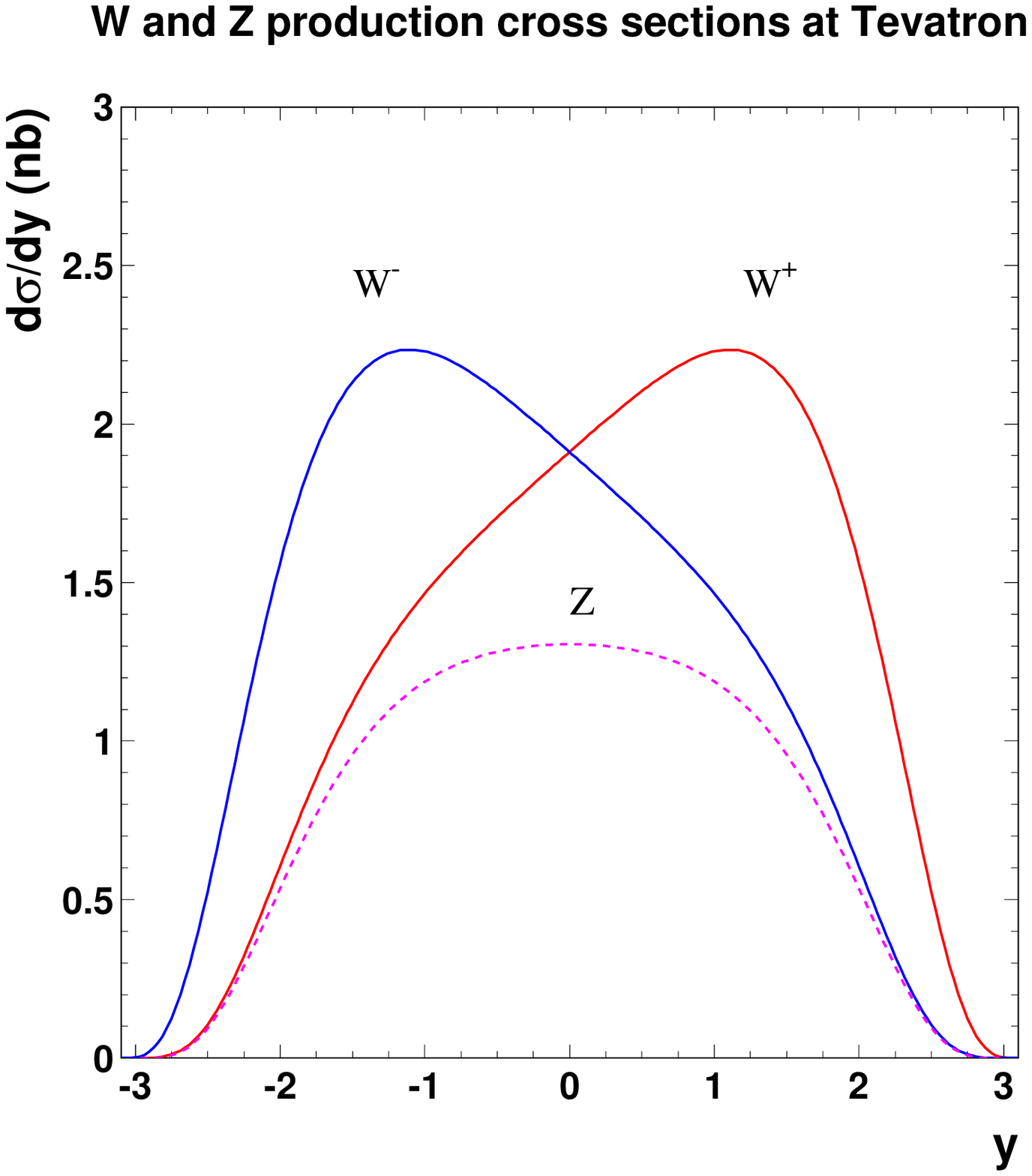}
\includegraphics[width=6cm]{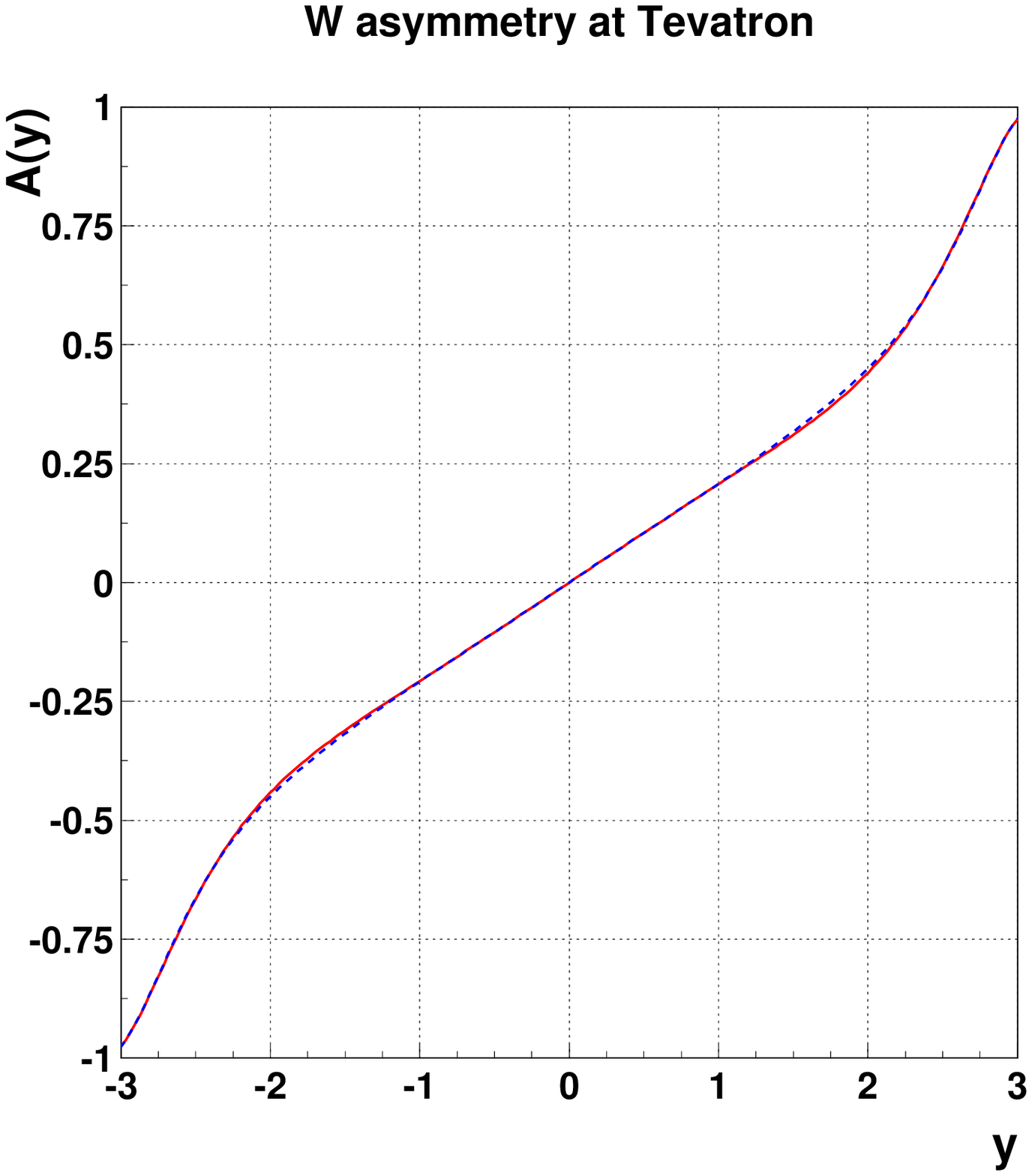}
\caption{Left: the $W$ and $Z$ boson production cross sections at Tevatron as functions of the boson rapidity $y$ for the MSTW08 parton distributions. Right: the $W$  boson asymmetry (solid line) together
with the approximate relation (\ref{eq:udapprox}) (dashed line).}
\label{fig:1}
\end{center}
\end{figure}

Assuming that the fraction $x_1$ refers to the proton and the fraction $x_2$ refers to the antiproton in the
$p\bar{p}$ scattering, the $W$ production cross sections are related to the nucleon parton distributions in the following way,
see Fig.~\ref{fig:0},
\beeq\nonumber
\frac{d\sigma_{W^+}}{dy} &\sim& u_p(x_1)\, \dbar_{\pbar}(x_2)\,+\,\dbar_p(x_1)\, u_{\pbar}(x_2)
\\\label{eq:ppbar1}
\frac{d\sigma_{W^-}}{dy} &\sim& d_p(x_1)\, \ubar_{\pbar}(x_2)\,+\,\ubar_p(x_1)\, d_{\pbar}(x_2)\,.
\eeeq 
From the charge conjugation symmetry we have
\be
d_{\pbar}(x)=\dbar_p(x)\,,~~~~u_{\pbar}(x)=\ubar_p(x)~~~~{\rm and}~~~~
\dbar_{\pbar}(x)=d_p(x)\,,~~~~\ubar_{\pbar}(x)=u_p(x)\,,
\ee
thus,  we find
\beeq\nonumber
\frac{d\sigma_{W^+}}{dy} &\sim& u_p(x_1)\, d_p(x_2)\,+\,\dbar_p(x_1)\, \ubar_{p}(x_2)
\\\label{eq:ppbar2}
\frac{d\sigma_{W^-}}{dy} &\sim& d_p(x_1)\, u_{p}(x_2)\,+\,\ubar_p(x_1)\, \dbar_{p}(x_2)\,.
\eeeq 
Notice that interchanging $x_1 \leftrightarrow x_2~ (y\to -y)$ we have: 
$d\sigma_{W^+}/{dy} \leftrightarrow d\sigma_{W^-}/{dy}$, and
\be\label{eq:sym1}
\frac{d\sigma_{W^+}(y)}{dy}=\frac{d\sigma_{W^-}(-y)}{dy}\,.
\ee
This is clearly seen in Fig.~\ref{fig:1} (left) where
the weak boson production cross sections are shown 
for the {proton-antiproton} collisions at the  Tevatron energy
$\sqrt{s}=1.8~{\rm TeV}$ (in which case $y_{max}\approx 3.1$). We use the LO MSTW08 parametrization \cite{Martin:2009iq} of the parton distribution functions.
Form relation (\ref{eq:sym1}), the $W$ boson production asymmetry $A(y)$ is odd function rapidity, see Fig.~\ref{fig:1} (right).

In  $p\pbar$ collisions, the $W$-charge asymmetry in rapidity is also defined
\be
A_{W^+}(y)=\frac{d\sigma_{W^+}(y)/dy-d\sigma_{W^+}(-y)/dy}{d\sigma_{W^+}(y)/dy+d\sigma_{W^+}(-y)/dy}\,.
\ee
Form the above mentioned symmetry, we have the following chain of equalities
\be
A_{W^+}(y)=A_{W^-}(-y)=A(y)\,.
\ee
These asymmetries are useful for the determination
of the parton distributions since assuming for simplicity the local isospin symmetry of the sea quark distributions, 
$
\dbar_p(x)=\ubar_p(x) 
$, 
we obtain from  relations (\ref{eq:ppbar2})
\be\label{eq:udexact}
A(y)=\frac{u_p(x_1)\, d_p(x_2)-d_p(x_1)\, u_{p}(x_2)}{u_p(x_1)\, d_p(x_2)+d_p(x_1)\, u_{p}(x_2)
+2\,\ubar_p(x_1)\,\ubar_p(x_2)}\,.
\ee
For most of the parton distribution  parametrizations, the term $2\,\ubar_p(x_1)\,\ubar_p(x_2)$ in the denominator
can be neglected and we find
\be\label{eq:udapprox}
A(y)\simeq \frac{u_p(x_1)\, d_p(x_2)-d_p(x_1)\, u_{p}(x_2)}{u_p(x_1)\, 
d_p(x_2)+d_p(x_1)\, u_{p}(x_2)}\,.
\ee
The quality of this relation is shown in  Fig.~\ref{fig:1} (right) where the solid curve shows Eq.~(\ref{eq:udexact}) 
while the dashed curve corresponds to  relation (\ref{eq:udapprox}), computed for the LO MSTW08 parametrisation.

Relation (\ref{eq:udapprox}) is the basis of the current analyses of the Tevatron data on the $W$ production asymmetry
for the determination of the ratio $d_p(x)/u_p(x)$, since from Eq.~(\ref{eq:udapprox}) we find
at the scale $\mu=M_W$
\be\label{eq:udapprox1}
\frac{d_p(x_1)/u_p(x_1)}{d_p(x_2)/u_p(x_2)}\simeq \frac{1-A(y)}{1+A(y)}\,.
\ee


\subsection{$pp$ collisions}

\begin{figure}[t]
\begin{center}
\includegraphics[width=6cm]{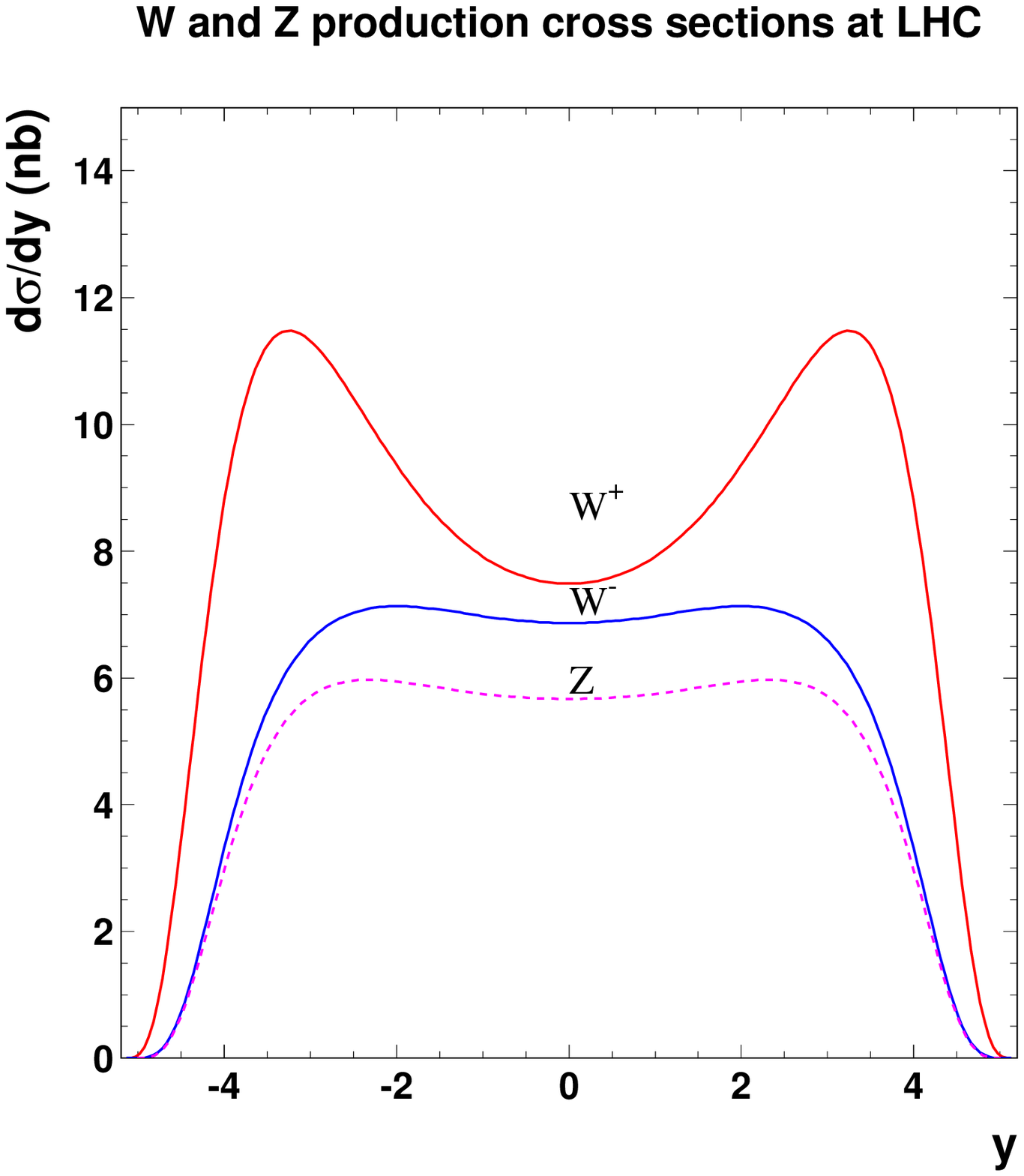}
\includegraphics[width=6cm]{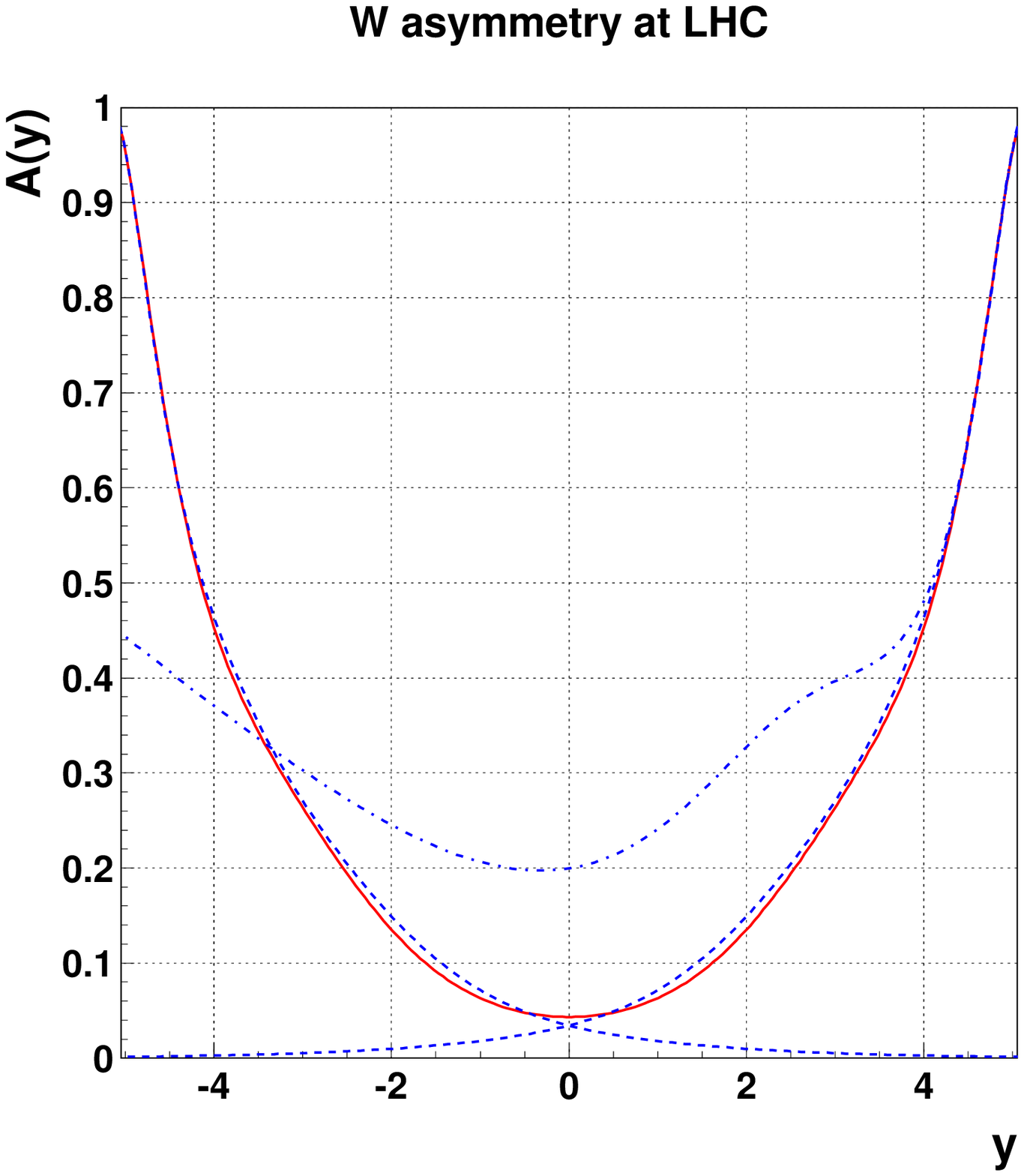}
\caption{Left: the $W$ and $Z$ boson production cross sections at the LHC as functions of the boson rapidity $y$ for the  LO MSTW08 parton distributions. Right: the $W$  boson asymmetry (solid curve) together with 
relation (\ref{eq:asympp}) computed   at $x=x_1$ and $x=x_2$ (two dashed lines) and for  the valence quark distributions at $x=x_1$ (dash-dotted  line).}
\label{fig:2}
\end{center}
\end{figure}

For $p{p}$ collisions, the $W$ production cross sections look as follows
\beeq\nonumber
\frac{d\sigma_{W^+}}{dy} &\sim& u_p(x_1)\, \dbar_{p}(x_2)\,+\,\dbar_p(x_1)\, u_{p}(x_2)
\\\label{eq:pp1}
\frac{d\sigma_{W^-}}{dy} &\sim& d_p(x_1)\, \ubar_{p}(x_2)\,+\,\ubar_p(x_1)\, d_{p}(x_2)\,.
\eeeq 
Due to  symmetric proton beams,
the transformation $x_1\leftrightarrow x_2$ leaves $d\sigma_{W^\pm}$ unchanged, which is
reflected in the symmetry of these cross sections under the rapidity reflection $y\to -y$.
This is clearly seen in Fig.~\ref{fig:2} (left) where the cross sections for  the LHC energy $\sqrt{s}=14~{\rm TeV}$ (in which case $y_{max}\approx 5.1$) are shown for
the LO MSTW08  parton distributions. Assuming for simplicity the local isospin symmetry  for the sea quark distributions, $\ubar_p(x)=\dbar_p(x)$, we find
\be\label{eq:asymppp}
A(y)=\frac{(u_{p}(x_1)-d_{p}(x_1))\,\ubar_{p}(x_2)\,+\, \ubar_{p}(x_1)\,(u_{p}(x_2)-d_{p}(x_2))}
{(u_{p}(x_1)+d_{p}(x_1))\,\ubar_{p}(x_2)\,+\, \ubar_{p}(x_1)\,(u_{p}(x_2)+d_{p}(x_2))}\,,
\ee
which is evidently even function of rapidity, see Fig.~\ref{fig:2} (right). 
In the limit $x_1\sim 1$ and $x_2\ll  1$ or $x_1\sim x_2\ll 1$, 
the sea quark distribution $\ubar(x_1)$ is small and the second terms in 
the numerator and denominator of Eq.~(\ref{eq:asymppp}) can be neglected. Thus, we obtain
\be\label{eq:asympp}
A(y)\simeq \frac{u_{p}(x_1)-d_{p}(x_1)}{u_{p}(x_1)+d_{p}(x_1)}\,.
\ee
From the $x_1\leftrightarrow x_2$ symmetry, the same relation holds true when
the argument of the parton distributions  in Eq.~(\ref{eq:asympp}) is changed to $x_2$. These approximate
relations are shown by the two dashed curves in Fig.~\ref{fig:2} (right). 
Thus, from the measurement of the $W$ asymmetry  in the $pp$ collisions,
the $d_{p}(x)/u_{p}(x)$ ratio at the scale  $\mu=M_W$ can be extracted,
\be\label{eq:dupp}
\frac{d_{p}(x)}{u_{p}(x)}\simeq \frac{1-A(y)}{1+A(y)}\,,
\ee
down to $x_1\simeq M_W/\sqrt{s}\approx 0.006$ for the LHC energy.
Relation (\ref{eq:asympp}) can also be
written in terms of the valence $(val)$ and sea $(sea)$ quark distributions 
\beeq\nonumber
u_p(x)\eq u_{val}(x)+u_{sea}(x)\,,~~~~~~~~~~~\ubar_p(x)=u_{sea}(x)
\\\label{eq:valsea}
d_p(x)\eq d_{val}(x)+d_{sea}(x)\,,~~~~~~~~~~~\dbar_p(x)=d_{sea}(x)\,,
\eeeq
taking the following form (assuming isospin symmetry)
\be\label{eq:asymppnew}
A(y)\simeq \frac{u_{val}(x_1)-d_{val}(x_1)}{u_{val}(x_1)+d_{val}(x_1)+(u_{sea}(x_1)+d_{sea}(x_1))}\,.
\ee
For $x_1\to 1$ the sea quark distributions in the denominator can be neglected and
relation (\ref{eq:dupp}) gives 
\be\label{eq:dupppp}
\frac{d_{val}(x)}{u_{val}(x)}\simeq \frac{1-A(y)}{1+A(y)}\,.
\ee
The quality of this relation for the LO MRTW09 parametrization is shown in Fig.~\ref{fig:2} by
the dash-dotted line  computed from eq.~(\ref{eq:asympp}) with the valence quark distributions
at $x=x_1$ (a symmetric curve can be found for $x=x_2$). We see that for $y>4$ ($x>0.3$) relation (\ref{eq:dupppp}) can  be    used for the determination of  the $d_{val}/u_{val}$ ratio at the scale $\mu=M_W$.

\section{Diffractive production of $W$  bosons}
\label{chapter:3}

\begin{figure}[t]
\begin{center}
\includegraphics[width=6cm]{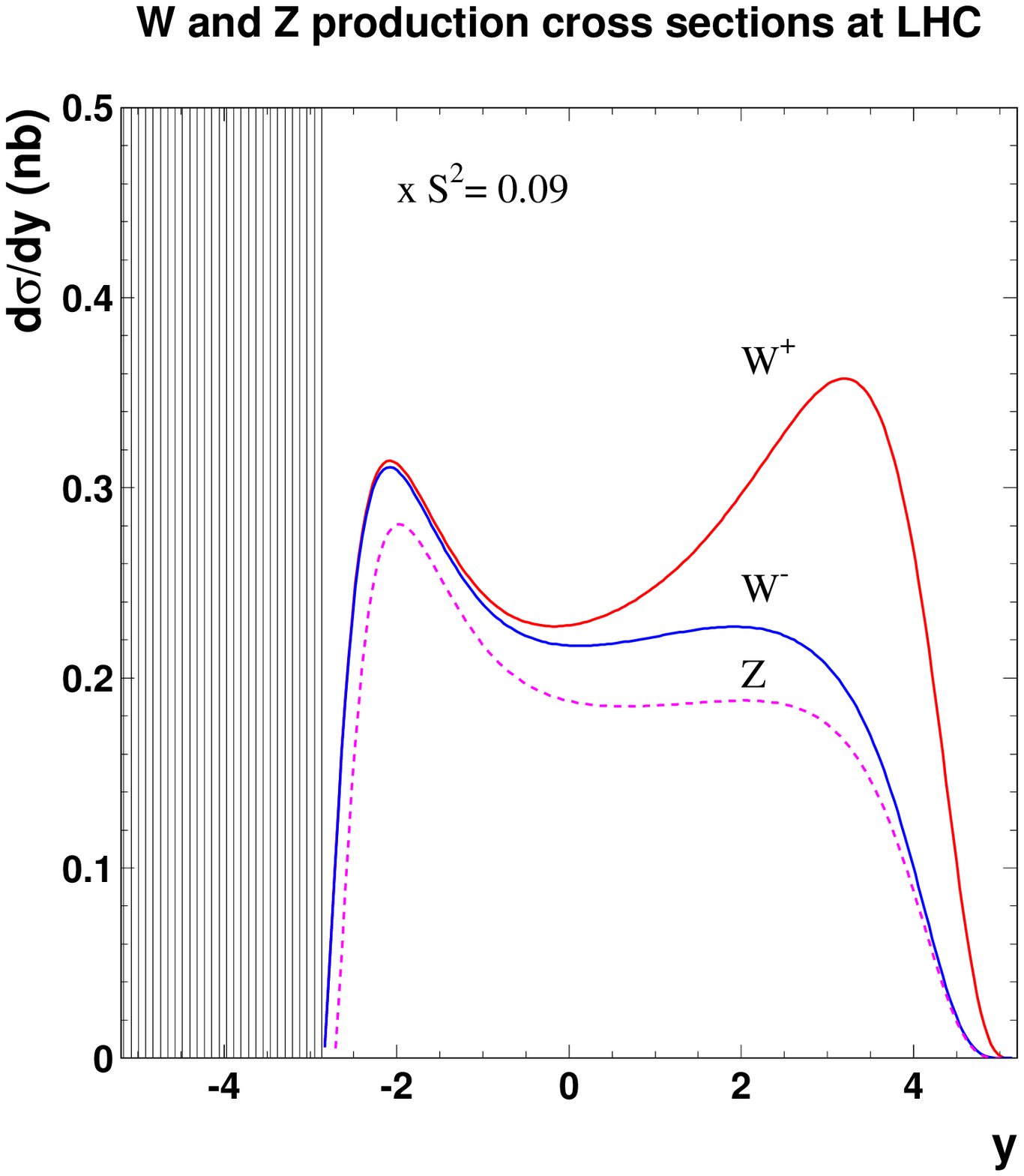}
\includegraphics[width=6cm]{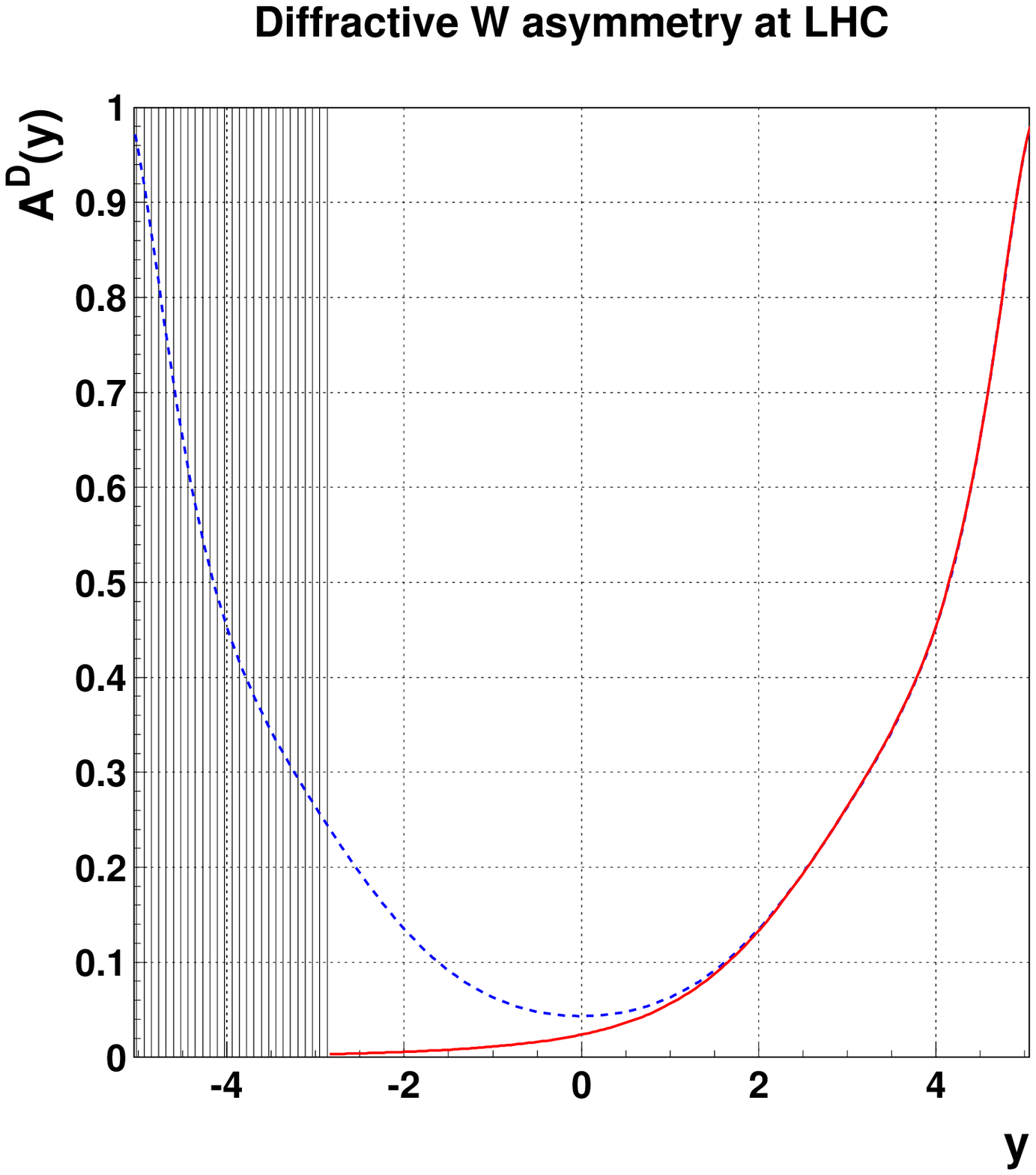}
\caption{Left: the single diffractive $W/Z$ boson production cross sections at the LHC as functions of boson rapidity. The results have to be multiplied by the gap survival factor $S^2=0.09$.
Right: the $W$  asymmetry  in $p\funp$ collisions (solid line), given by Eq.~(\ref{eq:asympom1}), together with the  asymmetry (\ref{eq:asymppp}) in  $pp$ collisions (dashed line). The shaded areas indicate  the rapidity gap $\Delta=2.3$
for $\xp=0.1$.
}
\label{fig:3}
\end{center}
\end{figure}

In the diffractive case, the electroweak bosons are produced in a restricted region of rapidity with a rapidity gap without particles between the proton which stayed intact and the diffractive system
from the dissociated proton. In this process boson mass is a hard scale allowing
for perturbative QCD interpretation as in the nondiffractive case.
However, the nature of the vacuum quantum number exchange, which leads to the rapidity gap, is nonperturbative. It
is usually modelled using the Regge theory notion - a pomeron. In the model of Ingelman and Schlein 
\cite{Ingelman:1984ns} the pomeron is endowed with a partonic structure described by the pomeron parton distributions $q_\funp$, which replace the standard inclusive parton distributions on the dissociated  proton side. Since the pomeron carries vacuum quantum numbers, these distributions have to be flavour symmetric 
\be\label{eq:pomeronsym}
u_{\funp}(x)=\bar{u}_{\funp}(x)=d_{\funp}(x)=\bar{d}_{\funp}(x)=s_{\funp}(x)=\bar{s}_{\funp}(x)=\ldots\equiv q_\funp(x)\,,
\ee 
where $x=x_2/\xp$ with $\xp={M^2_{D}}/{s}$ being a fraction of the proton's momentum transferred into the diffractive system of mass $M_{D}$. With such a definition, $x$ is a fraction of the pomeron
momentum carried by the parton taking part in the $W$ boson production. From the condition 
 $0<x, \xp<1$,  one finds that the   $W$ boson rapidity is in the range
\be\label{eq:gap}
-y_{max}+\ln(1/\xp)<y<y_{max}\,,
\ee
and the rapidity gap has the length  $\Delta=\ln(1/\xp)$.

Thus, in the single diffractive case, the $W$ production cross sections are related to quark distributions in the following way
\beeq
\frac{d\sigma_{W^+}}{dyd\xp} &\sim& (u_{p}(x_1)\,+\,\dbar_{p}(x_1))\,q_\funp(x_2/\xp)
\\\label{eq:ppom}
\frac{d\sigma_{W^-}}{dyd\xp} &\sim& (d_{p}(x_1)\,+\,\ubar_{p}(x_1))\,q_\funp(x_2/\xp)\,.
\eeeq 
In more general approach, the pomeron parton distribution should be replaced by diffractive parton 
distributions \cite{Berera:1994xh,Berera:1995fj,Trentadue:1993ka,Collins:1994zv,Collins:1997sr}, which in the pomeron model interpretation  have the Regge factorized form
\be 
q_D(x_2,\xp)=f(\xp)\,q_\funp(x_2/\xp)\,,
\ee 
where $f(\xp)$ is called pomeron flux. Independent of this interpretation, however, the diffractive quark  distributions should also be flavour symmetric. 
In Fig.~\ref{fig:3} (left) we show the $W$ and $Z$ production cross sections with the 
 LO MSTW08  proton parton distributions and the pomeron parton distributions from the analysis
\cite{GolecBiernat:2007kv}. The effect of the pomeron in the left hemisphere is clearly visible in the left hemisphere - the rapididty gap is formed and the $W^\pm$ asymmetry strongly decreases.
These  cross sections should be  multiplied by a gap survival factor, $S^2=0.09$ \cite{Khoze:2000wk}, which 
takes into account soft  interactions destroying the rapidity gap.

\begin{figure}[t]
\begin{center}
\includegraphics[width=6cm]{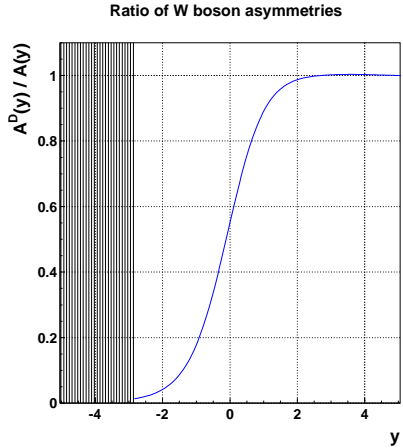}
\caption{The ratio of the $W$ boson production asymmetries in the diffractive and nondiffractive $pp$ scattering.
}
\label{fig:4}
\end{center}
\end{figure}

The $W$ boson production asymmetry (\ref{eq:6})  is a particularly good observable since it is
insensitive to  the gap  survival probability  \cite{Bjorken:1992er}
which  multiplies both the  cross sections $d\sigma_{W^{\pm}}/dyd\xp$.
The flavour symmetric pomeron parton distributions also  cancel, and we obtain
for the $W$ asymmetry in the diffractive case,
\be\label{eq:asympom1}
A^{D}(y)=\frac{u_{p}(x_1)-d_{p}(x_1)\,+\,\dbar_{p}(x_1)-\ubar_{p}(x_1)}{u_{p}(x_1)+d_{p}(x_1)
\,+\,\dbar_{p}(x_1)+\ubar_{p}(x_1)}\,,
\ee
where  the parton distributions are taken at the scale $\mu=M_W$. Notice that $A^D(y)$ is independent of
$\xp$, i.e. the length of the rapidity gap.
Substituting decomposition
(\ref{eq:valsea}), we find
\be\label{eq:asympom2}
A^{D}(y)=\frac{u_{val}(x_1)-d_{val}(x_1)}
              {u_{val}(x_1)+d_{val}(x_1)\,+\,2\,(u_{sea}(x_1)+d_{sea}(x_1))}\,.
\ee
This is an exact result obtained  only under the assumption (\ref{eq:pomeronsym}).
In  Fig.~\ref{fig:3} (right) we show the asymmetry (\ref{eq:asympom2}) (solid line) together  with 
the $W$ boson asymmetry (\ref{eq:asymppp}) in the inclusive case (dashed line).

In order to understand  our result,  it is interesting to compare 
Eq.~(\ref{eq:asympom2}) with the approximate asymmetry (\ref{eq:asymppnew}),
valid in the right hemisphere for $y>0$. For large rapidities, when the sea quark distributions can be neglected, these two asymmetries are equal while for $y\approx 0$, when the valence quark distributions
in the denominator are negligible, $A^D(y)\approx A(y)/2$. This is clearly seen in  Fig.~\ref{fig:4} where the ratio
$A^D(y)/A(y)$, with $A(y)$ given by Eq.~(\ref{eq:asymppp}), is shown. Approaching the rapidity gap, the
asymmetry $A^D(y)\to 0$ 
while $A(y)$ rises. Thus,  the ratio shown  in Fig.~\ref{fig:4} is close to zero
at the edge of the rapidity gap.

The pattern shown in Fig.~\ref{fig:4} is quite general and depends only on the assumption on flavour symmetry of the pomeron  parton distributions, Eq.~(\ref{eq:pomeronsym}). Therefore, it would be  interesting to test experimentally the very concept of the flavour  symmetric pomeron parton distributions  by measuring the ratio 
of the two $W$ asymmetries in the diffractive and nondiffractive $pp$ scattering. Systematic errors will cancel in such a ratio which  should allow for   quite precise determination of this quantity.
We are looking forward to the  experimental verification of the presented results at the LHC.

\section{Summary}

The measurement of the $W$ boson production asymmetry in the diffractive $pp$ collisions is a valuable method to test the concept of  the flavour symmetric  pomeron parton distributions. If it is true, the
$W$ asymmetry in the single diffractive case provides an additional constraint for the parton distribution
functions in the proton.

\begin{acknowledgments}
Useful discussions with Ewelina \L{}obodzinska, Christophe Royon and Jacek Turnau are gratefully acknowledged.
This work is partially supported by the grants of MNiSW Nos.  N N202 246635 and
N N202 249235, and the grant HEPTOOLS, MRTN-CT-2006-035505.
\end{acknowledgments}

\bibliographystyle{h-physrev4}
\bibliography{mybib}

\end{document}